\documentclass[11pt]{article}
\usepackage{amssymb,cite}
\newcommand\blank[1]{}

\newcommand\toline[1]{--#1}

\newcommand{\fract}[2]{{\textstyle\frac{#1}{#2}}}

\newcommand{\ri}{\right}
\newcommand{\alp}{\alpha}

\newcommand{\lf}{\left}
\newcommand{\te}{\theta}

\newcommand{\CL}{{\cal L}}

\newcommand\NN{{\mathbb N}}

\newcommand\ZN{{\mathbb Z}_N}

\newcommand{\balpha}{\alpha\kern -6.7pt\alpha}
\newcommand{\bbalpha}{\alpha\kern -4.95pt\alpha}

\newcommand{\CM}{{\cal M}}

\newcommand\eq{\begin{equation}}
\newcommand\en{\end{equation}}
\newcommand\bea{\begin{eqnarray}}
\newcommand\eea{\end{eqnarray}}
\newcommand\nn{\nonumber}

\newcommand{\eff}{\rm eff}
\newcommand{\One}{{\hbox{{\rm 1{\hbox to 1.5pt{\hss\rm1}}}}}}
\renewcommand{\One}{{\mathbb 1}}
\renewcommand{\One}{{\rm 1\!\!1}}

\newcommand{\ceff}{ c_{\rm eff}}
\newenvironment{tab}{\linespread{1.0} \begin{table}}{\end{table}%
\linespread{1.3}}
\begin{document}
\vskip 0.5cm
\begin{center}
{\Large \bf Finite size effects and the supersymmetric
  sine-Gordon models} 
\end{center}
\vskip 0.8cm
\centerline{Clare Dunning%
\footnote{\tt tcd1@york.ac.uk}
}
\centerline{\sl\small 
Dept.~of Mathematics, University of York, York YO1 5DD, UK }
\vskip 0.9cm
\begin{abstract}
\vskip0.15cm
\noindent
We propose  nonlinear integral equations to describe the
groundstate energy of the 
fractional supersymmetric sine-Gordon models. The equations  encompass
the $N=1$ supersymmetric sine-Gordon model as well as the 
$\phi_{\rm id,id,adj}$ perturbation of the $SU(2)_L \times SU(2)_K /
SU(2)_{L{+}K}$ models at rational level $K$. 
A second set 
of equations are 
proposed for the groundstate energy of  the $N=2$ supersymmetric
sine-Gordon model.

\end{abstract}
\setcounter{footnote}{0}
\def\thefootnote{\fnsymbol{footnote}}

\section{Introduction}
\label{intro}
The sine-Gordon (SG) and supersymmetric sine-Gordon (SSG)
 \cite{DS,JH,FGS,GS,SY} models 
 belong to an interesting set of 
integrable quantum field theories that are said to have fractional
supersymmetry \cite{BL,Zamf}, thus attracting the  name fractional
 supersymmetric 
sine-Gordon models (FSSG).  The field content  \cite{ABL,BL2} consists
of a single 
boson interacting with  a $Z_L$ generating parafermion,  the choices
 $L=1$ and $2$
yielding  the SG and SSG models respectively. 
The models correspond to the fractional conformal field  theory obtained from
the Wess-Zumino-Witten model $SU(2)_L$ by compactifying the free
 boson  on a circle of  
radius $2R=\beta/\sqrt{4\pi}$,  perturbed by 
\eq
\Phi(z ,\bar z) = \frac{g \sqrt {4\pi}}{\beta} \Psi_1(z) \bar \Psi_1(z)
  e^{-i \frac{\beta}{\sqrt 4\pi}\varphi(z,\bar z)} +{\rm  c.c.}~.
\en
 The mass scale is set by the dimensionful parameter $g$, 
 and the real dimensionless parameter 
$\beta$ determines the particle content.
Since the boson $\varphi(z ,\bar z)$ 
and  generating parafermion $\Psi_1(z ,\bar z)$   
have conformal dimensions $1$ and 
 $1{-}1/L$ respectively, the 
perturbing term has conformal dimension  $\beta^2/8\pi {+}1{-}1/L$.  The 
 ultraviolet effective central charge is $1{+}2(L{-}1)/(L{+}2)=
 3L/L{+}2$. 

In the limit $\beta^2/8\pi \to 1/L$ 
the perturbed model becomes the current-current perturbation of the WZW
 model $SU(2)_L$ \cite{ABL}. Instead if we tune  the parameter  
\eq
\xi=\frac{L\beta^2/8\pi}{1/L-\beta^2/8\pi}
\label{xibeta}
\en 
to $K{+}2$ we obtain the 
unitary  coset $SU(2)_L \times SU(2)_K /SU(2)_{L+K}$ 
perturbed by the  operator $\phi_{\rm id,id,adj}$ \cite{ABL}. 
The
indices indicate  which representation 
of  $SU(2)_L, SU(2)_K$ and 
$SU(2)_{L{+}K}$ is associated to the corresponding weights. 
This choice of $\xi$ when $L=1$ 
corresponds to the  well-known quantum group restriction
\cite{SmsG,lecl}  of the 
 sine-Gordon model  
to the   $\phi_{13}$ perturbed minimal models. 
In fact, the  FSSG models were first proposed as the models obtained by 
 `unrestricting'  the  S-matrices describing the  $\phi_{\rm
 id,id,adj}$ perturbation of the cosets
 $\CM_{L,K}\equiv SU(2)_L \times SU(2)_K /SU(2)_{L+K}$\cite{ABL}.  

The existence of higher spin conserved charges in  1+1 dimensional
integrable quantum field theories strongly constrain
the allowed scattering processes.      There is 
 no particle production, the individual particle momenta are 
conserved, and all 
 $n$-particle scattering processes decompose into 
products of two-particle scattering amplitudes~\cite{ZZSmat}.
The S-matrix of a 
supersymmetric model should  also 
commute with the supersymmetry generators. 
This typically leads to a further factorisation of the two-particle
S-matrix  into a direct product in 
which one  factor encodes  the 
supersymmetry dependence \cite{ABL,Sch,BL2}.
The  S-matrices of the FSSG models exactly fit this pattern \cite{ABL,BL2}: 
\eq
S(\beta) = S_{ L} \otimes S_{\rm SG}(\beta)~.
\label{sfac}
\en 
The supersymmetry-related factor 
 $S_{ L}$ 
describes  the scattering of the massive $\phi_{13}$ 
perturbation of the minimal model $\CM_{1,L}$ \cite{Zamf,lecl,BL}, and  
the bosonic  factor
$S_{\rm SG}(\beta)$ is the S-matrix of the sine-Gordon  model
at coupling $\beta$ \cite{ZZSmat} . The particle content
of the FSSG consists of  a soliton of mass $M$ and its 
antisoliton  together with a number of breathers of mass 
\eq 
M_j=2M \sin\lf (\frac{\pi j\,\xi}{2}\ri) \quad,\quad j=1\dots<\frac{1}{\xi}~.
\en
There are no breathers in the  repulsive regime $\xi<1$. 
Once the spectrum and associated  S-matrices have been conjectured
the thermodynamic Bethe ansatz \cite{yang,ZamTBA} is usually  
employed to check the proposed scattering theory. However when the
scattering is nondiagonal, as above, deriving  the TBA equations from
the S-matrices  becomes rather difficult.

Fendley and Intriligator noted that one can immediately write
down a TBA system for  
a model with a tensor product 
S-matrix of the form $S_G \otimes S_H$  if the 
 TBA equations are already known for the  models described
 by S-matrices 
$S_G$ and $S_H$ \cite{FI}. If the individual TBA
equations are encoded on Dynkin-like diagrams of type $G$ and $H$, each
having a single massive node, 
then the TBA
equations for the  model with S-matrix  $S_G \otimes S_H$
are found by 
gluing the individual TBA equations at the massive node 
\cite{FI}.  
In the compact notation of \cite{RTV1,RTV2}
this corresponds to $(G \diamond H)_L$, where $L$ indicates the
massive node. 
It is easy to check that 
 the TBA equations found for the FSSG model 
at the special values of the coupling \cite{FI,FZa,RTV1,RTson,RT}
\eq
\frac{\beta^2}{8\pi} = \frac{1}{L} - \frac{2}{m} \quad , \quad
m \in \NN^+ \quad,\quad  L=1,2,\dots \le {\rm int}[m/2]{-}1 ~,
\label{betaxi}
\en
and  those for the  $\phi_{\rm id,id,adj}$  perturbation of
the $\CM_{L,K}$ models \cite{Zab} all fit this pattern.    TBA equations  for 
the models arising as the 
$\Phi_{13}$ perturbation of the 
supersymmetric  models ${\cal S}\CM_{2,4n+4}$
 are known \cite{AhnYL,FSn2,MelzerSUSY,MS}, but currently lack an 
interpretation in terms of Dynkin-like diagrams. 

A $Y$-system, from 
which the TBA equations can be obtained via Fourier transform
\cite{Zamy},  for the 
FSSG model at any coupling has been proposed using the gluing idea and
the sine-Gordon $Y$-system \cite{RTson,RT,KSS}. The SG $Y$-system
depends on writing the smaller of $\xi$ or $1/\xi$  
as a simple continued fraction. The 
number of TBA equations is found to be equal to the sum
of the leading terms of the denominators in the continued fraction
plus one further equation. 
Though there is an neat
depiction in terms of Dynkin diagrams \cite{RTson,RT} it is hard to
deal with 
a large number of equations. Moreover those
$\xi$ for which 
 the continued fraction does not  terminate 
 are associated to  a  TBA system with an infinite number of equations.

There is an alternative method of calculating the finite size effects
of the sine-Gordon model, known as the nonlinear
integral 
equation (NLIE) 
technique. The result is a single equation, first proposed for the
groundstate of the sine-Gordon model  by 
 Destri and De~Vega in 1992~\cite{DDV}. A complete derivation to 
include the excited states of even topological charge followed 
in \cite{FRT1}, and an equation for the excited states of odd
topological charge was proposed in \cite{FRT4}.  
The 
 conformal limit of the groundstate equation had been proposed in 1991
 in an  
integrable lattice model context by Kl\"umper, Batchelor and
Pearce~\cite{PK,KBP}. Further motivation for this work  
came from the the lattice model connection: the $XXX$ spin chain  
at spin $S$ is a model with underlying field theory $SU(2)_{2S}$. The
spin-$\frac{1}{2}$ case corresponds to the 
lattice model mentioned 
above; the spin-$1$ problem is discussed in~\cite{KBP,PK}.     
Recently
Suzuki has generalised the technique of \cite{PK,KBP} to obtain a 
a finite set of nonlinear integral equations that  describe the
thermodynamics of the spin-$S$ chain at finite temperature 
\cite{Suzspin}.  Given the relations between the spin chains and
$SU(2)_{2S}$, and the FSSG models and perturbed $SU(2)_L$, 
we should expect the conformal limit of any integral equation
describing the FSSG 
at $\beta^2=1/L$ to be closely related to the zero temperature version of
Suzuki's NLIE. 

Motivated by Suzuki's equations and the gluing idea 
we propose  a set of nonlinear
integral equations to   describe the groundstate energy of the
 fractional 
supersymmetric sine-Gordon model at any value of the coupling constant
$\beta^2$.  In \S2 a short description of the
component equations leads to the proposed NLIEs, and then a 
 number of tests are carried out in \S3.

The $N{=}2$ supersymmetric sine-Gordon model is
not a member of the FSSG series. However the groundstate energy of
this model can be studied by means of a further set of NLIEs, obtained
by a simple generalisation of the FSSG equations. 
The details are in \S4 along
with a short discussion of related models: the
sausage models and the parafermionic perturbation of the $Z_M$ models. 
Directions for future work can be found in \S5.

\section{Nonlinear integral equations}
There are two  integrable quantum field theories associated
with the $\phi_{13}$ perturbation of the minimal model $\CM_{1,L}$. 
If the coupling $\lambda$ is positive the action 
\eq
{\cal A}={\cal A}_{\rm CFT}  + \lambda \int d^2 x \, \phi_{13}
\label{act}
\en
describes a `massless flow', or renormalisation group trajectory, from 
$\CM_{1,L}$ in the ultraviolet to $\CM_{1,L{-}1}$ in the infrared \cite{Za}. 
On the other hand, if the coupling is negative (\ref{act}) describes
a massive 
integrable quantum field theory with S-matrix $S_L$ \cite{Zaint}.  
 The TBA equations for the groundstate energy of the  massive
 scattering theory were found by Al.B.\ Zamolodchikov \cite{ZRSOS}. As we
 explain below  one small change allows the same equations to unexpectedly 
describe the groundstate energy of the massless flows \cite{Zc}.
The equations are written  in terms
of  the incidence matrix 
$I_{ab}$ of the $A_{L}$ Dynkin diagram \cite{ZRSOS,Zc}. 
The TBA system is\footnote{Note that the TBA equations  are usually written in
  terms of  
   pseudoenergy $ \varepsilon=-f$.}  
\bea
f_a(\te) &=& g_a(\te) +\sum_{b=1}^{L}
 \int_{-\infty}^\infty  d\te'\,I_{ab}\,\chi(\te-\te')  \ln(1+e^{f_b(\te')}) \nn \\
c_{\eff}(r) &=& \frac{6}{\pi} \sum_{a=1}^{L} 
\int_{-\infty}^{\infty} d\te\, g_a(\te) \ln(1+e^{f_a(\te)})~,
\eea
with kernel
$
\chi(\te)=1/(2\pi\cosh \te) .
$
Here $r=MR$ where $M$ is  the mass of the elementary kink, or in the
massless theory the crossover scale,  and $R$ is the
 circumference of the cylinder on which the model is considered. 
The massive theory is found by setting the driving term to 
\eq
g_a (\te)= - \delta_{a,1}\, r \cosh\ \te 
\en
while the massless theory appears if 
\eq
g_a (\te)= - \frac{r}{2} \,(\delta_{a,L}  \, e^{ \te} +  \delta_{a,1}
\,e^{-\te})~. 
\en
Both theories have the same 
  ultraviolet effective central charge: 
 $\ceff(0)=1{-}6/(L{+}2)(L{+}3)$, while the infrared value is zero or 
$ 1{-}6/(L{+}1)(L{+}2)$ for the massive/massless models respectively.

To avoid  dealing with a different and possibly infinite set of
 equations for each choice of $\beta^2$ we consider the second
 type of 
 nonlinear integral equation mentioned in the introduction.  
The equation 
involves a complex function $a(\te)$ and
its  conjugate $\bar a(\te)$ coupled via \cite{PK,KBP,DDV}
\bea
\ln a(\te)&=&\!\!\!\!-i\,r\sinh \te  +\int_{{\cal C}_1} 
\!d\te'\,\varphi(\te{-}\te')\ln (1+a(\te')) -
\int_{{\cal C}_2} 
\!d\te'\,\varphi(\te{-}\te')\ln (1+a^{-1}(\te'))~.
\label{sgnlie}
\eea
The integration contours 
${\cal C}_1$ and ${\cal C}_2$ run from $-\infty$ 
to $+\infty$, just below and just above the real $\theta$-axis
respectively. 
The 
kernel $\varphi(\te)$ is   proportional to the logarithmic
derivative of the
soliton-soliton scattering amplitude of the sine-Gordon model
\eq
\varphi(\te)=\int \frac{dk}{2\pi}\,e^{i k\te}
\frac{\sinh (\xi{-}1)\frac{\pi k}{2}}
{2\sinh \frac{\pi\,\xi\, k}{2} \cosh \frac{\pi k}{2} }~.
\label{sgker}
\en
The effective central charge found using 
\eq
c_{\rm eff}(r)= \frac{3ir}{\pi^2}
\left( \int_{{\cal C}_1}d\te\, \sinh \te \, \ln
  (1+a(\te)) 
-\int_{{\cal C}_2}d\te\, \sinh \te \,\ln
(1+a^{-1}(\te))\right)
\en
has ultraviolet value  $1$.  With constant $i\pi\alpha$ 
\cite{Zd} added to
the RHS of (\ref{sgnlie}) the effective central charge becomes 
$\ceff(0) = 1-6 \alp^2 \xi /(\xi+1)$. 
The choice $\xi = p/(q{-}p)$ and $\alpha =\pm
1/p$ \cite{Zamp,FMQR,FRT3} implements the restriction \cite{SmsG,lecl} of
the SG model to 
the $\phi_{13}$ perturbed  minimal models.  

The explanation for Fendley and Intriligator's gluing proposal lies in
the derivation of the TBA equations \cite{FI}.  
The key is the factorisability of the S-matrix: it implies  
each factor can be 
diagonalised separately and each generates an individual set of
pseudoparticles. These 
are seen to only couple through the massive particle, with the 
 result being the  `factorised' or glued TBA system described in the
introduction. 
There is no equivalent of the pseudoparticles in the derivation  
of the sine-Gordon NLIE. Moreover, 
in further contrast to the TBA, the S-matrix plays no role in the NLIE
derivation. Instead 
one notes a posteriori that the kernel function of the
resulting NLIE is related as described above to the  
 sine-Gordon S-matrix. It therefore seemed unlikely that the gluing
 idea would  be of  any use in finding a NLIE for a model with 
 an S-matrix of the form (\ref{sfac}).  Fortunately this turns out not
 to be the case. 

We obtain the  NLIEs for the FSSG model by gluing 
the minimal model TBA equations  to the sine-Gordon NLIE at the
massive node following 
the Fendley-Intriligator prescription \cite{FI}. However the resulting
equations are not quite right and  it is necessary 
 to make a 
few simple modifications.  Due to singularities in 
$\ln(1+y_1(\te))$ for real $\te$ a TBA-like convolution $\int d\te'
\chi(\te-\te') 
\ln(1+y_1(\te'))$ along the real axis cannot be used. The 
solution is to replace this term with a convolution  along 
the contours ${\cal C}_1$ and ${\cal 
  C}_2$ encircling the real axis as in the SG NLIE. 
To ensure for real $\te$ the reality of the $L{-}1$  TBA-like
 functions $y_2,\dots,y_L$, and the complex nature of the NLIE-like
 function $y_1$ we find we should  couple 
$y_1$ and $y_2$ 
to each other with an extra shift of $i\pi/2$ in the  kernel
$\chi$. The conjectured equations are 
\bea
\ln y_1(\te) &=& \nu_1(\te) +i \pi \alpha \nn \\
&&\quad +\int_{{\cal C}_1} 
\!d\te'\,\varphi(\te{-}\te')\ln (1+y_1(\te')) -
\int_{{\cal C}_2} 
\!d\te'\,\varphi(\te{-}\te')\ln (1+y_1^{-1}(\te'))
\nn \\ 
&&\quad \quad +\int_{-\infty}^{\infty} 
\!d\te'\,\chi(\te{-}\te'{+}\fract{i\pi}{2})\ln (1+y_2(\te'))~;
\nn \\  [11pt]
\ln y_2(\te) &=& \nu_2(\te)+\int_{{\cal C}_1} 
\!d\te'\,\chi(\te{-}\te'{-}\fract{i\pi}{2})\ln (1+y_1(\te')) -
\int_{{\cal C}_2} 
\!d\te'\,\chi(\te{-}\te'{-}\fract{i\pi}{2})\ln (1+y_1^{-1}(\te'))
\nn \\  &&
\quad \quad +\int_{-\infty}^{\infty} 
\!d\te'\,\chi(\te{-}\te')\ln(1+y_3(\te'))~;
  \nn  \\ 
\ln y_a (\te)&=&  \nu_a(\te)+\sum_{b=1}^{L} I_{ab}\int_{-\infty}^{\infty}\!d\te'
\, \chi(\te-\te') \ln(1+y_b(\te')) 
\quad , \quad a=3 \dots  L ~,\label{nlieqs}
\eea
where  $\chi(\te)$ remains as above.  The  driving terms are 
\eq
\nu_a(\te)= \left\{ 
\begin{array}{cl}
 -i  r \sinh (\te)  & \quad a=1 \\ 
0 & \quad {\rm otherwise}
\end{array} \right. ~.
\label{my}
\en
Borrowing the TBA notation of \cite{RTV1,RTV2} our shorthand for 
equations of this type will be $(A_1 \diamond A_L)_a^{[i]}$. The
superscript $[i]$ indicates the NLIE-like 
function and the subscript $a$ the function on which the mass is
placed. The equations  above are therefore represented by  $(A_1
\diamond A_L)^{[1]}_1$.  
The effective central charge  defined in terms of the 
solution of  (\ref{nlieqs}) can be calculated using 
\eq
c_{\eff}(r) = -\frac{3}{\pi^2} \sum_{a=1}^L \ \int_{{\cal C}_1} d\te\,
\nu_a (\te) \ln(1+y_a(\te)) -   \int_{{\cal C}_2} d\te \,
\nu_a (\te) \ln(1+y_a^{-1}(\te))~.
\label{csg}
\en
At $L=1$ the system (\ref{nlieqs},\ref{my},\ref{csg}) trivially
reduces to the 
NLIE describing the  
sine-Gordon equation. As expected these equations become equivalent to 
Suzuki's \cite{Suzspin} for the spin-$S$  $XXX$  spin chain in the limit
$\beta^2 \to 1/L$ ($\xi\to\infty$) . The NLIE-like functions are
approximately related
via  $y_1 (\te) 
\approx \ln
\mathfrak{a}(\fract{2}{\pi}\te+i)$, provided 
the spin chain driving term is replaced with the one relevant for the
field theory.

Based on previous experience with  NLIEs 
we expect to find the effective central charge for  the FSSG model at
level $L$ and coupling 
$\beta^2$  by tuning $\xi$ according to (\ref{xibeta}) and
setting $\alpha=0$. The  $\phi_{\rm id,id,adj}$ perturbation of the
coset $\CM_{L,K}$ should be 
found at 
\eq 
\xi=K{+}2 \quad,\quad \alpha=\pm 1/(K{+}2)~.
\label{xicoset}
\en  
In fact we hope  the equations go a little further: even though the
quantum group restriction of the S-matrices may not work it is
thought that the 
FSSG models also describe the nonunitary cosets perturbed by $\phi_{\rm
id,id,adj}$. If we set 
\eq
\xi=\frac{Lp}{q{-}p} \quad,\quad \alpha =\pm \frac{1}{p}~,
\label{xinon}
\en 
then the NLIEs should yield  the effective central charge of the
$\phi_{\rm   id,id,adj}$ perturbation 
of  $\CM_{L,Lp/(q{-}p){-}2}$. This is  a coset at  
rational, and possibly negative, level. These  models make sense
provided the   integers 
$p,q$ are such that $p$ and $(q-p)/L$ are coprime integers. At $L=1$ this is
the $\phi_{13}$ perturbation of the minimal models $\CM(p,q)$, and at
$L=2$  the supersymmetry preserving perturbation $\Phi_{13}$ of the
superminimal models ${\cal S}{\cal M}(p,q)$.

In the same way that TBA equations for  the   
massive perturbation of the minimal models can be converted into those
describing  the 
massless perturbations, we can describe 
{\it some} massless flows using  the proposed NLIEs with new  
  driving terms 
\eq
\nu_a(\te)= \left\{ 
\begin{array}{ll}
   -\fract{i}{2} r e^{ \te}  & \quad a=1 \\ [6pt]
 - \fract{1}{2}r e^{-\te}\, \delta_{a, \,L{-}K{+}1} & \quad {\rm otherwise}
\end{array} \right.~. 
\label{ly}
\en
If  $(\xi,\alpha)$  are fixed to the values prescribed
in (\ref{xicoset}) the equations reproduce the flow
 $\CM_{L,K}+\phi_{\rm id,id,adj} \to 
\CM_{L,K{-}L}$. 
Given the different nature of $y_1$ and $y_{L{-}K{+}1}$
it is rather surprising that the resulting NLIEs do in fact describe 
massless flows. 
The massless flows within the sine-Gordon model ($L=1$) can
alternatively be
studied via 
  two coupled NLIE functions \cite{Zd,DDT}.

Using the symmetry between $L$ and $K$ we can minimise the number of
NLIEs for the coset models $\CM_{L,K}$ by choosing $L<K$. However the
massless flows only make sense if $L>K$. 
Both possibilities lead to a 
smaller number of equations than the relevant TBA equations. 
Therefore the new NLIEs
are a  
computationally more efficient way  of calculating the finite size
effects of the FSSG models.  
In the next section we take a closer at the equations and obtain some
results confirming the above conjectures. 

\section{Testing  the NLIEs}

We begin by
examining the behaviour of the individual functions at $r=0$ and
$r=\infty$. By differentiating the first equation of (\ref{nlieqs})
with respect to 
$\te$ we see the function $i\ln(y_1)(\te)$ 
becomes approximately constant in the region $|\te| \ll \ln(2/r)$ as
$r \to 0$. 
For $\te \ll -\ln (2/r)$ it behaves as $-e^{r\te}$, and for $\te \gg
\ln(2/r)$ as $e^{r\te}$. Therefore $y_1$ is a  complex constant,
which we denote $x_1^0$,   inside the 
the central region $ -\ln(2/r) \ll\te\ll \ln(2/r)$. 
The real and imaginary parts of $y_1$ oscillate outside this region
but because $\eta$ is nonzero $y_1(\te-i\eta)$ and
$y_1^{-1}(\te+i\eta)$    ultimately tend to  zero.  
The functions  $y_2,y_3,\dots ,y_L$ 
behave exactly as if they were TBA Y-functions\footnote{in standard notation 
$Y_a(\te)=\exp[\varepsilon_a(\te)]$}: they are symmetric in $\te$  and
real for real $\te$.  
From the behaviour of $y_1$ it follows that in the 
ultraviolet limit they become  constants  $x_a^{-} , x_a^0
, x_a^{+}$ in the regions
$\te \ll -\ln (2/r), -\ln (2/r) \ll \te \ll \ln (2/r) $ and $\te
\gg \ln (2/r)$ respectively, while  interpolating smoothly in between. 
A similar conclusion can be reached for the  stationary solutions of
the massless equations, apart from a small change in the behaviour of  
$y_1$ as it now  becomes a nonzero constant in the region $\te \ll
 -\ln (2/r)$.  It is simple to check this picture by plotting 
 some numerical solutions to the NLIEs. 
The exact values of the stationary
 solutions, derived from the NLIEs in the usual way \cite{ZamTBA,KM1},
 are summarised for both massive and  massless cases in table \ref{tabM}. 
%
%
\begin{tab}
\begin{center}
\begin{tabular}{l| c|c |  c | c  }
 &  a & $1+x_a^- $ & $1+x_a^0$  & $1+x_a^+$ \\ \hline 
\rule[0.5cm]{0cm}{2mm} 
& $1$ & $1$   & $ \frac{\sin \frac{\pi  \xi \, \alp (L{+}1)
  }{L{+}\xi}}{\sin\frac{\pi \xi \, \alp}{L{+}\xi}}\, e^{i\pi \xi\,\alp
L/(L{+}\xi)}   $ & $ 1   $ \\ 
\rule[0.5cm]{0cm}{2mm} 
\raisebox{3ex}[0cm][0cm]{Massive}& $2\dots L$ &   $\frac{\sin^2 \frac{\pi a}{L{+}2}}
{\sin^2 \frac{\pi}{L{+}2}}  $  & $\frac{\sin^2 \frac{\pi  \xi \, \alp (L{+}2{-}a)
  }{L{+}\xi} }{\sin^2 \frac{\pi \xi \, \alp}{L{+}\xi}}  $  & $\frac{\sin^2 \frac{\pi a}{L{+}2}}
{\sin^2 \frac{\pi}{L{+}2}} $   \\ \hline
\rule[0.5cm]{0cm}{2mm} 
& $1$ &  $\frac{ \sin\frac{\pi (K{-}L)}{L{+}2}}{\sin\frac{\pi}{L{+}2}}
\, e^{i \pi(K{+}1)/(L{+}2)}  $   & $ \frac{\sin \frac{\pi (L{+}1) 
  }{L{+}K{+}2}}{\sin\frac{\pi }{L{+}K{+}2}}\, e^{i\pi 
L/(L{+}K{+}2)}   $ & $ 1   $ \\ 
\rule[0.5cm]{0cm}{2mm} 
Massless &$2\dots L{-}K$ &   $ \frac{\sin^2 \frac{\pi(K{+}a)}{L{+}2}}
{\sin^2 \frac{\pi}{L{+}2}} $  & $\frac{\sin^2 \frac{\pi  (L{+}2{-}a)
  }{L{+}K{+}2} }{\sin^2 \frac{\pi }{L{+}K{+}2}}  $  & $\frac{\sin^2
  \frac{\pi a}{L{+}2}} 
{\sin^2 \frac{\pi}{L{+}2}} $  \\
\rule[0.5cm]{0cm}{2mm} 
& $L{-}K{+}1 \dots L$ &   $ \frac{\sin^2 \frac{\pi(L{+}2{-}a)}{K{+}2}}
{\sin^2 \frac{\pi}{K{+}2}} $  & $\frac{\sin^2 \frac{\pi  (L{+}2{-}a)
  }{L{+}K{+}2} }{\sin^2 \frac{\pi }{L{+}K{+}2}}  $  & $\frac{\sin^2
  \frac{\pi a}{L{+}2}} 
{\sin^2 \frac{\pi}{L{+}2}} $  
\end{tabular}
\caption{  \protect{Stationary values for massive and massless
    perturbed models.}
\label{tabM}}
\end{center}
\end{tab}

Using the 
 standard methods \cite{ZamTBA,KM1,KBP} we find the ultraviolet 
 effective central charge  takes the form 
\bea
c_{\eff}(0) &= & -\frac{3}{\pi^2} \Bigl[
2\CL\lf(\frac{1}{1+\bar x_1^0}\ri)-\CL\lf(\frac{1}{1+\bar x_1^{+}}\ri) 
-\CL\lf(\frac{1}{1+\bar x_1^{-}}\ri) \nn \\ 
&& \quad+ \sum_{a=1}^{L} 
2\CL\lf (\frac{1}{1+x_a^0} \ri) - 
 \CL\lf(\frac{1}{1+x_a^+}\ri)- \CL\lf(\frac{1}{1+x_a^-}\ri)  \nn \\ &&
 \quad \quad
-\frac{i\pi \alp}{2}  \Bigl ( \ln \frac{ (1+x_1^0)^2}{(1+x_1^-)(1+x_1^+)}
-\ln \frac{ (1+\bar x_1^0)^2}{(1+\bar x_1^-)(1+\bar x_1^+)} \Bigr) \Bigr]~,
\label{cth}
\eea
where $\CL$ is Rogers dilogarithm function 
\eq
\CL (x)=
-\frac{1}{2} \int_{0}^ x  dt\, \frac{\ln 1-t}{t} + \frac{\ln t}{1-t}~.
\en
To evaluate (\ref{cth}) we use the standard properties and
known sum rules  of the dilogarithm (see for example
  \cite{Kir})   
\eq
\frac{6}{\pi^2} \sum_{a=2}^{L} \CL\lf ( \frac{\sin ^2[ \pi/(L+2)]}
{\sin ^2[\pi a/(L+2)]}  \ri) = \frac{2(L-1)}{L+2} \quad,\quad
\frac {6}{\pi^2}\CL(1)=1~,
\en
and one  
 further  sum rule which involves both real and  complex arguments 
\eq
\frac{6}{\pi^2} \lf( \CL\lf(\frac{1}{1+\bar x_1^{0}}\ri)+ \sum_{a=1}^L\CL \lf
(\frac{1}{1+x_a^{0}}\ri)\ri) =1~.
\label{sum}
\en
Numerical calculations confirm that (\ref{sum}) holds for {\it arbitrary}
real numbers $\xi$ and $\alpha$, $L$ is by definition an integer. 
The ultraviolet effective central charge predicted from the NLIEs
takes  the form 
\eq
c_{\eff}(0) =  \frac{3L}{L+2} - \frac{6\alp^2\xi L}{L+\xi}~.
\label{cfin}
\en
Does this reproduce the expected results? Yes, it does: 
with  $\xi$ related to $\beta^2$ as in (\ref{xibeta}) and $\alpha=0$,
the formula predicts 
$\ceff(0)=3L/(L{+}2)$, whereas the choice (\ref{xicoset}) leads to
the effective central charge of the unitary coset models $\CM_{L,K}$: 
\eq
c_{\eff}(0) =  \frac{3L}{L+2} - \frac{6\, L}{(L+K+2)(K+2)}~,
\label{ccos}
\en
and the possibility (\ref{xinon}) yields the effective central charge 
for the nonunitary coset models 
\eq
\ceff(r)=\frac{3L}{L+2} - \frac{6L}{pq}~.
\en 

If we use the massless driving terms, the individual parts of 
(\ref{cth}) combine in a 
different way, but the final result is again (\ref{cfin}). Since we
have fixed the 
parameters $\xi=K+2$ and $\alp=\pm 1 /(K{+}2)$ we 
actually obtain (\ref{ccos}). We need to check the infrared
limit to be sure the NLIEs really do describe these massless flows. 
As $R\to \infty$ the contribution to $\ceff(\infty)$ from
$y_{L{+}\xi+3},\dots,y_{L}$ becomes 
vanishingly small. The remaining functions 
have stationary solutions given by the massless ones in  table~\ref{tabM}
but with $L$  
replaced by  
 $L{-}K$. Plugging these values into (\ref{cth}) gives 
\eq
\ceff(\infty) =  \frac{3L}{L+2} - \frac{6\, L}{(K+2)(K-L+2)}~,
\en
which is exactly the ultraviolet effective central charge of $\CM_{L ,K{-}L}$. 

Before obtaining the conformal dimension of the perturbing operator
we comment on a surprising relation between the NLIEs at the coset
points and the associated TBA equations.
The massive  $\phi_{\rm id,id,adj}$ perturbation of $\CM_{L,K}$
has a TBA system 
 based on $(A_1 \diamond A_{L{+}K{-}1})_L$. The symmetry between 
$L$ and $K$ means  that we can either take a NLIE with $L$ functions
(below denoted $y_a$)
and $\xi=K{+}2$, or one  with $K$ functions (denoted  $ \tilde y_a$) and
$\xi=L{+}2$, in either case $\ceff(r)$ is exactly the same.
Curiously the magnonic parts of both NLIEs are closely related to the 
 TBA functions. Numerical evidence indicates
\eq
 \exp( f_a) = \left\{
\begin{array}{cl}
y_{L{-}a{+}1} &   \quad a=1,\dots,L{-}1 \\ 
\tilde y_{a{-}K{+}1}&  \quad a=L{+}1 , \dots , L{+}K{-}1  
\end{array} \right.~.
\en
The functions with nonzero driving term, $\exp f_L, y_{1}
,\tilde y_{1}$, are not equal, but the integral equations imply 
they must be related. One such relation is 
\bea
\quad  \int_{-\infty}^\infty  d\te'\, \chi(\te-\te')
 \ln(1+e^{f_{L{-}2}(\te')}) &=&  \nn \\ 
 && \hspace{-4.6cm}\int_{{\cal C}_1} 
\!d\te'\,\chi(\te{-}\te'{-}\fract{i\pi}{2})\ln (1+y_1(\te')) -
\int_{{\cal C}_2} 
\!d\te'\,\chi(\te{-}\te'{-}\fract{i\pi}{2})\ln (1+y_1^{-1}(\te'))~.
\eea

The question of the conformal dimension of the
perturbing operator, predicted as usual from the periodicity of the
functions entering the NLIEs \cite{Zamy}, can be partially answered
via the relation  described above with the coset TBA systems.
 For $\xi=K{+}2$ and $\alpha=\pm 1/(K{+}2)$ the periodicity of the
 TBA-like functions  
 $y_2 , \dots, y_L$ inferred from that of the TBA systems for the
 $\phi_{\rm id,id,adj}$ perturbation of the coset $\CM_{L,K}$
  is $i\pi (\xi+L)$ \cite{Zab}. 
 It is not expected to depend on $\alpha$ and
 so one only needs to check what happens at other values of $\xi$.
Doing so numerically leads us 
to believe  all of the entire
functions  $y_a(\te)$ have periodicity $i\pi (\xi+L)$, and therefore 
a Laurent 
expansion in powers of $\exp(2/(\xi+L)\te  )$. 
The quantities 
$\ln(1+y_a(\te))$ appearing in the kernels of the NLIEs have
essentially the same shape as the 
quantities $\ln(1+\exp(-\varepsilon_a(\te))$ that appear in a pure TBA system, 
that is a central plateau which quickly goes to zero for $\te$
outside the region $|\te| \ll \ln(2/r)$. Therefore we may employ
Zamolodchikov's  argument \cite{Zamy} to  predict that 
$\ceff(r)$ expands in powers of $r^{4/(\xi+L)}$. Note that for some
$\alpha$,  $x_1^{0}=0$ and Zamolodchikov's argument acquires a small 
modification, as explained in\cite{DTc}, with the  result that
$\ceff(r)$ instead expands in 
powers of $r^{2/(\xi+L)}$. However for the choices of $\alpha$ needed
here $x_1^0$ is never identically zero. 
If the perturbing operator is
odd we would expect an expansion in 
$r^{2(1-\Delta)}$ leading to the prediction $\Delta = 1-2/(\xi+L)$. 
Inserting the appropriate value of $\xi$  (\ref{xinon}) we 
find  
$\Delta = 1-2/(K{+}L{+}2),
$
which exactly matches the conformal dimension of $\phi_{\rm id,id,adj}$ for
both unitary 
and nonunitary (with $K=Lp/(q-p)$) cosets. On the other hand if the
perturbing operator is even we would expect an expansion in
$r^{4(1-\Delta)}$ which implies $\Delta = 1{-}1/(\xi{+}L)$. Writing $\xi$
in terms of $\beta^2$ using (\ref{xibeta}) we find the conformal
dimension  of the FSSG models: $\Delta = \beta^2 
/8\pi{+}1{-}1/L$. 

In the infrared limit the groundstate energy of the massive model
expands in powers of $e^{-MR}$ \cite{ZamTBA}.  Adapting the 
arguments of \cite{ZamTBA,DDV} we obtain from 
the NLIEs 
\eq
E(R) \sim -4 \cos \frac{\pi}{L{+}2}  \cos \pi \alpha
\int\frac{d\te}{2\pi}\,\cosh\te \,e^{-MR \cosh \te}~.
\en
With the choices of $L$ 
and $\alpha$ as given in the text this agrees  perfectly 
with TBA results obtained for 
the perturbed 
cosets $\CM_{L,K}+\phi_{\rm id,id,adj}$ \cite{Zab}, the $\Phi_{13}$
perturbation of the 
$N=2$ supersymmetric minimal models \cite{FI} and the $SU(2)$ Gross-Neveu
model \cite{PF2}.

We now test the NLIEs numerically, comparing against the effective
 central charge found using various
 TBA equations~\cite{Zab,RTson,FI,RTV1}.  These TBA 
equations correspond either to  $\CM_{L,K}+\phi_{\rm id,id,adj}$ for
 integer $K$, or to the 
FSSG model at the specific coupling given in (\ref{betaxi}). We 
 also compare against 
the magnonic 
TBA system based on $ (A_1 \diamond T_n)_L$, identified
in \cite{RTV1} with
$\CM_{L,n{-}3/2}+\phi_{\rm  id,id,adj}$. However  this is only true
at $L=1$ and for
general $L$ these TBA equations correspond to 
 $\CM_{L,n{-}L{-}1/2}+\phi_{\rm id,id,adj}$. 
As no  data for the  chosen cases 
has previously been reported in the literature 
we solved both the TBA and NLIE equations numerically, using an
 iterative method. We  checked 
 stability  of the solution by varying the number of iterations and the 
 integration step size.  
The results are
 shown in  
table~\ref{tabmart}. The tables include a diagrammatic
version of the TBA system with a solid circle representing the
massive node. We also 
reproduce one massless flow whose TBA diagram has crosses to indicate the
nodes on which  the massless driving terms are placed.
In all cases the agreement is very good. 
\begin{tab}[ht]
\begin{center}
\begin{tabular}{l |c|  c| l| c |c }
 &   & NLIE &  & TBA&  NLIE
\\ ~~~~~~\raisebox{1.5ex}[0cm][0cm]{Model}
& \raisebox{1.5ex}[0cm][0cm]{TBA}&  $(L,\xi,\alpha)$ &~~~\raisebox{1.5ex}[0cm][0cm]{$r$} &   $\ceff(r)$ & $
\ceff(r)$
\\ \hline  
\rule[0.2cm]{0cm}{2mm} 
 &  \setlength{\unitlength}{.8mm}
\begin{picture}(20,6)(0,2)
\linethickness{.4pt} 
\put(1,1){\circle{2}}
\put(2,1){\line(1,0){2}}
\put(5,1){\circle{2}}
\put(6,1){\line(1,0){2}}
\put(3.8,.35){\tiny $+$}
\put(11.8,.35){\tiny $+$}
\put(9,1){\circle{2}}
\put(10,1){\line(1,0){2}}
\put(13,1){\circle{2}}
\put(14.2,1){\line(1,0){1.7}}
\put(17,1){\circle{2}}
\end{picture}   && 1E-05   &  1.2499968080510   & 1.2499968080550
   \\ 
 \raisebox{1.5ex}[0cm][0cm]{$\CM_{2,4} {+}\phi_{\rm id,id,adj}
$  }& & 
 \raisebox{1.5ex}[0cm][0cm]{$(4,4,\pm\frac{1}{4})$} &100&
1.0002668224816 & 1.0002668224816
\\\hline

 & \setlength{\unitlength}{.8mm}
\begin{picture}(20,6)(2.,2)
\linethickness{.4pt}
\put(1,1){\circle{2}}
\put(2,1){\line(1,0){2}}
\put(5,1){\circle{2}}
\put(6,1){\line(1,0){2}}
\put(9,1){\circle*{2}}
\put(10,1){\line(1,0){2}}
\put(13,1){\circle{2}}
\put(14.2,1){\line(1,0){1.7}}
\put(17,1){\circle{2}}
\qbezier[40](18,1) (21,4)(22,1)
\qbezier[40](18,1) (21,-2)(22,1)
\end{picture}     & &  0.01  & 1.5789577694079  &  1.5789577694082
   \\
 \raisebox{1.5ex}[0cm][0cm]{$\CM_{3,\frac{3}{2}} {+}\phi_{\rm id,id,adj}
$   }&&    \raisebox{1.5ex}[0cm][0cm]{$(3,\frac{7}{2},\pm\frac{1}{7})$
 }  &  0.3  &1.3491110222344 & 
 1.3491110222443 \\
\hline
 &   \setlength{\unitlength}{.8mm}
\begin{picture}(20,6)(2.,2)
\linethickness{.4pt}
\put(1,1){\circle{2}}
\put(2,1){\line(1,0){2}}
\put(5,1){\circle{2}}
\put(6,1){\line(1,0){2}}
\put(9,1){\circle*{2}}
\put(10,1){\line(1,0){2}}
\put(13,1){\circle{2}}
\put(14.2,1){\line(1,0){1.7}}
\put(17,1){\circle{2}}
\put(18,1){\line(1,0){2}}
\put(21,1){\circle{2}}
\end{picture}     &  & 0.01 & 1.4590240568136 &  
1.4590240568137    \\ 
\raisebox{1.5ex}[0cm][0cm]{$\CM_{3,4} {+}\phi_{\rm id,id,adj}$}
&&  \raisebox{1.5ex}[0cm][0cm]{$(3,6,\pm\frac{1}{6})$}  & 0.3&   1.2821840541083 & 1.2821840541085 \\ \hline
\rule[0.2cm]{0cm}{2mm} 
   &
  \setlength{\unitlength}{.8mm}
\begin{picture}(11,6)(1,2)
\linethickness{.4pt}
\put(1,1){\circle{2}}
\put(2.2,1){\line(1,0){2}}
\put(5,1){\circle*{2}}
\put(6,1){\line(1,0){2}}
\put(9,1){\circle{2}}
\put(9,2){\line(0,1){1.8}}
\put(9,5){\circle{2}}
\put(9,-.2){\line(0,-1){1.8}}
\put(9,-3){\circle{2}}
\end{picture}  
& & 0.01 &   
1.4889948011771 & 1.4889948011793  \\  
  \raisebox{1.5ex}[0cm][0cm]{$S_{2} \otimes S_{\rm SG} (\frac{3}{10})
    $} & &  \raisebox{1.5ex}[0cm][0cm]{$ (2,3,0)$}   & 
 0.3 & 1.2975198090220 &
1.2975198092226  \\   
\hline
\rule[0.2cm]{0cm}{2mm} 
  &
\setlength{\unitlength}{.8mm}
\begin{picture}(11,6)(1,2)
\linethickness{.4pt}
\put(1,1){\circle{2}}
\put(2.2,1){\line(1,0){1.8}}
\put(5,1){\circle{2}}
\put(6,1){\line(1,0){2}}
\put(9,1){\circle{2}}
\put(9,2){\line(0,1){3}}
\put(9,5){\circle*{2}}
\put(9,-.2){\line(0,-1){3}}
\put(9,-3){\circle*{2}}
\end{picture}   & & 0.01 & 1.9854196180142 & 1.9854196180142 \\
 \raisebox{1.5ex}[0cm][0cm]{$S_{4} \otimes S_{\rm SG}(\frac{1}{20})$
 } & &   \raisebox{1.5ex}[0cm][0cm]{$ (4,1,0)$}  &0.3 &
 1.7223919114950 &  
 1.7223919114948 
\end{tabular}
\caption{  \protect{
Comparison of 
$c_{\rm eff}(r)$ 
calculated using the TBA equations and the NLIE.}
\label{tabmart}}
\end{center}
\end{tab}

We can also enquire to which  models  the NLIEs
based on  $(A_1 \diamond A_L)^{[1]}_j$ correspond. If we set the 
 $j^{\rm th}$ driving term to 
$ -r\cosh\te$ and all others  to zero, the 
ultraviolet  effective central charge becomes
\eq
\ceff(0)= \frac{3(L-j+1)}{L-j+3} - \frac{6 \xi \alp^2
  (L-j+1)}{(\xi+L)(\xi+j-1)}~.
\en
For   $\xi{=}K{+}3{-}j$ and $\alp{=}\pm 1/\xi$ we rediscover  
 the massive  $\phi_{\rm id,id,adj}$ perturbation of a familiar model:
 $\CM_{L{-}j{+}1,K}$.

\section{$N{=}2$ supersymmetric sine-Gordon model}
Recent interest in the $N{=}2$ supersymmetric sine-Gordon model 
\cite{KU1} has arisen in two different contexts, both
related to string theory. Though little is currently known about the
excited state 
spectrum of the 
$N{=}2$ 
sine-Gordon model it is predicted to  provide nontrivial information on the
 spectrum of states of closed superstrings in certain
 backgrounds \cite{MM}.  Also, boundary versions of the
 $N=2$ SSG model  (see for example \cite{War,Nep})
are relevant for open string theory. 

The S-matrix of the $N{=}2$ SSG model has the usual
factorised form \cite{KU2,BL2,FMVW,FI}
\eq
S^{N{=}2}_{SG}(\tilde \beta) = S^{N{=}2} \otimes S_{\rm SG}(\beta)~.
\label{sneq2}
\en
The supersymmetric piece $ S^{N{=}2}$ is the 
sine-Gordon
S-matrix at its $N{=}2$ 
supersymmetric point $\beta^2/8\pi 
=2/3$. The 
 bosonic piece is 
the sine-Gordon S-matrix at coupling $\beta$. The $N{=}2$ coupling
constant $\tilde \beta$  is 
  related   to the sine-Gordon coupling via \cite{KU2}
\eq
\tilde \beta^2 = \beta^2 /(1-\beta^2/8\pi)
\en

TBA equations derived from the $S^{N{=}2}$ S-matrix 
 are encoded on $ (A_1 \diamond
A_3 )_2$ \cite{FI}. As noted in \cite{FSn2}, the TBA
equations for the $N{=}2$ SSG model can, in theory, be obtained by   gluing
the $(A_1 
\diamond A_3)_2$  
TBA system to that of the sine-Gordon model at coupling
$\beta$. In practice this has been done only for $\tilde \beta^2/8\pi
= k+1$, since at these points the SG TBA  reduces to  a simple Dynkin
diagram: $D_{k+2}$.   A second set of  simple TBA equations 
exist for some related models: the
 minimal $N=2$ conformal field theories perturbed by the superfield
 $\Phi_{13}$ \cite{FI}.

To describe the groundstate energy of
the $N{=}2$ SSG model at any coupling 
we replace the
complicated sine-Gordon TBA equations  with a single nonlinear
integral equation, as we did above for the FSSG models.    
We propose  equations similar to 
(\ref{nlieqs},\ref{my},\ref{csg}) but of type $(A_1 \diamond
D_3)^{[1]}_1$ with 
$\alpha=0$,  
rather than 
$(A_1 \diamond A_L)_1^{[1]}$. The nodes of $D_n$ are labelled such
 that $1$ is the first node of the tail, and $n,n{-}1$ label the fork
 nodes.The 
 stationary values 
$x_a^0$ are 
all infinite, which suggests that numerical solutions for very small
$r$ will converge extremely slowly, if at all (this is equally true for any
TBA equations with 
infinite stationary values). However combined with $x_2^\pm =x_3^\pm =1$  and
$x_1^\pm=0$  we find the correct ultraviolet central charge of
$3$. The leading logarithmic corrections to this value have been
proposed for some of the
models in \cite{Fatss}.  It is simple to check the proposed NLIEs 
have the expected  infrared limit.

If the mass is moved onto either of the magnonic nodes  so that the
NLIEs are of  type $(A_1 \diamond
D_3)^{[1]}_2$, we find 
the  effective
central charge of the sausage models
 $SST^{+}_{\lambda}$ \cite{FOZ}. 
Again the possibly
infinite set of TBA equations are in principle 
known.
The simplest, occurring at  $\lambda=1/M$, are  
represented by an extended $D_M$ Dynkin diagram \cite{FOZ}. A match is
found with the NLIEs if we set $\xi=M{-}2$ and $\alpha=0$. 
Alternatively if the twist $\alpha$ is
 set to
$\pm 1/(M{-}2)$  we find the $\psi_1\bar \psi_1 +\psi_1^\dagger  \bar
\psi_1^\dagger $ perturbation of 
the  $Z_{M{-}2}$  models, usually denoted $H^{(0)}_{M{-}2}$ \cite{F,FZa}. The
models denoted  $H^{(\pi)}_{M{-}2}$ refer to the massless 
perturbation of the $Z_{M{-}2}$ models. The flow of the
effective central charge from these models to the minimal models
$\CM(M{-}1,M)$ 
can be reproduced within the NLIEs provided 
the magnonic nodes are given driving terms of the form $ -
\fract{1}{2}r e^{\te} $ and $ - \fract{1}{2}r e^{-\te}$. Upon
resetting 
$\alpha$ to zero we find the effective central charge of the
`massless' sausage models $SST^{-}_{1/M}$.

Table~\ref{tabneq2} 
shows a numerical comparison of results found using this second set of
 NLIEs against some of the simplest TBA systems.   
\begin{tab}[ht]
\begin{center}
\begin{tabular}{l |c|  c| l| c |c }
  &  & NLIE &  & TBA&  NLIE
\\ 
~~~~~\raisebox{1.5ex}[0cm][0cm]{Model}
&\raisebox{1.5ex}[0cm][0cm]{TBA} &  $(\xi,\alpha)$ &~~~ 
 \raisebox{1.5ex}[0cm][0cm]{$r$} &   $\ceff(r)$ & $ 
\ceff(r)$
\\ \hline  
\rule[0.2cm]{0cm}{2mm} 

&   \setlength{\unitlength}{.8mm}
\begin{picture}(11,6)(2.2,2)
\linethickness{.4pt}
\put(5,1){\circle*{2}}
\put(5,-.1){\line(0,-1){1.8}}
\put(5,-3){\circle{2}}
\put(5,5){\circle{2}}
\put(5,2){\line(0,1){1.8}}
\put(6,1){\line(1,0){2}}
\put(9,1){\circle{2}}
\put(9,2){\line(0,1){1.8}}
\put(9,5){\circle{2}}
\put(9,-.2){\line(0,-1){1.8}}
\put(9,-3){\circle{2}}
\end{picture}
 && 0.01   &  2.5646432734143   &  2.5646432734187
   \\ 
 \raisebox{1.5ex}[0cm][0cm]{$S^{N{=}2}\otimes S_{SG}(\frac{3}{4})$}& & 
  \raisebox{1.5ex}[0cm][0cm]{$(3,0)$} &0.3&
 1.9298961954447 & 1.9298961954452
\\\hline

&   \setlength{\unitlength}{.8mm}
\begin{picture}(11,6)(2,2)
\linethickness{.4pt}
\put(7,1){\circle{2}}
\put(7,-.2){\line(0,-1){1.8}}
\put(7,-3){\circle*{2}}
\put(7,5){\circle{2}}
\put(7,2){\line(0,1){1.8}}
\put(8.1,1){\line(1,0){1.8}}
\put(11,1){\circle{2}}
\put(3,1){\circle{2}}
\put(4,1){\line(1,0){2}}

\end{picture}
 && 0.01   &  1.7401548125701   &  1.7401548125703
   \\ 
 \raisebox{1.5ex}[0cm][0cm]{$SST^{+}_{1/4} $}& &   
  \raisebox{1.5ex}[0cm][0cm]{$ (2,0)$} &0.3& 1.3902395891608
   & 1.3902395891609
\\\hline 
&
\setlength{\unitlength}{.8mm}
\begin{picture}(11,6)(0,2)
\linethickness{.4pt}
\put(1,1){\circle{2}}
\put(2.15,1){\line(1,0){1.8}}
\put(5,1){\circle{2}}
\put(6.2,1){\line(1,0){1.8}}
\put(9,1){\circle{2}}
\put(9,2){\line(0,1){3}}
\put(9,5){\circle*{2}}
\put(9,-.2){\line(0,-1){1.7}}
\put(9,-3){\circle{2}}
\end{picture}   & & 0.01 & 1.1373080354277 & 1.1373080354327 \\
 \raisebox{1.5ex}[0cm][0cm]{$H_5^{(0)}$
 } & &   \raisebox{1.5ex}[0cm][0cm]{$
(5,\pm\fract{1}{5})
$}  &0.3 & 1.0190548066399
 & 1.0190548066674
\\\hline 
&
\setlength{\unitlength}{.8mm}
\begin{picture}(11,6)(2,2)
\linethickness{.4pt}
\put(7.75,4.3){\tiny $+$}
\put(7.75,-3.7){\tiny $+$}
\put(5,1){\circle{2}}
\put(6,1){\line(1,0){1.8}}
\put(9,1){\circle{2}}
\put(9,2){\line(0,1){2}}
\put(9,5){\circle{2}}
\put(9,-.2){\line(0,-1){1.8}}
\put(9,-3){\circle{2}}
\end{picture}   & & 1E-05 & 0.9999980062005 & 0.9999980062002 \\
 \raisebox{1.5ex}[0cm][0cm]{$H_4^{(\pi)}$
 } & &   \raisebox{1.5ex}[0cm][0cm]{$
(4,\pm\fract{1}{4})
$}  &100 & 0.8003485312633
 &  0.8003485312633
 
\end{tabular}
\caption{  \protect{
Comparison of 
$c_{\rm eff}(r)$ 
found using  TBA equations and NLIEs of type $(A_1\diamond
D_3)_{a}^{[1]}$.}
\label{tabneq2}}
\end{center}
\end{tab}

\section{Summary}
The SS model  \cite{Fatss} is 
 a two-parameter family of integrable models whose name follows from the 
 S-matrix, which is  
a direct product of two sine-Gordon S-matrices. All of the models
 discussed above can be found as an SS model at specific values of
 the parameters \cite{Fatss,FB}. We leave for future investigation the 
 possibility of gluing two sine-Gordon NLIEs together to describe
 the groundstate energy of this family of models.

Motivated by the existing results for the sine-Gordon model
both sets of new NLIE 
should be generalised to  encode the full finite size
spectrum of the associated models.  Furthermore, it is curious that
the TBA driving 
term trick to find equations for the massless flows 
also works for the NLIEs; though only for the unitary coset models. 
A  rather different method  was
used in \cite{Zd} for the sine-Gordon model, also \cite{DDT,TCDw},
to obtain 
NLIEs describing massless flows at any value of $\beta^2$. It
remains to be seen whether a 
similar idea can be used on the new NLIEs. Finally it would be extremely
interesting to develop the finite size effects of the fractional
sine-Gordon models defined on a half-line.

\vspace{0.4cm}
\noindent{\bf Acknowledgments --} 
I'm grateful to Zoltan  Bajnok,  Andreas Kl\"umper, 
Paul Pearce, Francesco Ravanini and Gerard Watts for
useful discussions, and to Patrick Dorey and Roberto
Tateo for useful suggestions and discussions.
I thank the EPSRC for a Postdoctoral Fellowship and the Department of
Mathematics and Statistics, University of Melbourne for hospitality. 


\begin{thebibliography}{999}
\raggedright
\parskip 1pt
%

\bibitem{DS}
P.\ Di Vecchia and S.\ Ferrara,
{`Classical solutions in two-dimensional supersymmetric field
  theories'},
Nucl. Phys. B130 (1977) 93
%
\bibitem{JH}
J.\ Hruby,
{`On the supersymmetric sine-Gordon model and a two dimensional
  `bag''},
Nucl. Phys. B131 (1977) 275
%
\bibitem{FGS}
S.\ Ferrara, L.\ Giradello and S.\ Sciuto,
{`An infinite set of conservations laws of the supersymmetric
  sine-Gordon theory'},
Phys. Lett. B76 (1978) 303\toline{306}
%
\bibitem{GS}
L.\ Giradello and S.\ Sciuto,
{`Inverse scattering-like problem for supersymmetric models'},
Phys. Lett. B77 (1978) 267
%
\bibitem{SY}
I.\ Yamanaka and R.\ Sasaki,
{`Supervirasoro algebra and solvable supersymmetric quantum field
  theories'},
Prog. Theory. Phys. 79 (1988) 1167
%
\bibitem{BL}
D.\ Bernard and A.\ LeClair,
{`Residual quantum symmetries of the restricted sine-Gordon theories'},
Nucl. Phys. B340 (1990) 721\toline{751}
%
\bibitem{Zamf}
A.B.\ Zamolodchikov,
{`Fractional-spin integrals of motion in perturbed conformal field theory'},
Beijing 1989, Proceedings, Field, string and quantum
gravity. 349\toline{372}
%
\bibitem{ABL}
C.\ Ahn, D.\ Bernard and A.\ LeClair,
{`Fractional supersymmetries in perturbed coset CFTs and integrable  soliton theory'},
Nucl. Phys. B346 (1990) 409\toline{439}
%
\bibitem{BL2}
D.\ Bernard and A.\ LeClair,
{`The fractional supersymmetric sine-Gordon models'},
Phys. Lett. B247 (1990) 309\toline{316}
%
\bibitem{SmsG}
F.A.\ Smirnov,
{`Reductions of the sine-Gordon model as a perturbation of minimal models of conformal field theory'},
Nucl. Phys. B337 (1990) 156\toline{180}
%
\bibitem{lecl}
A.\ LeClair,
{`Restricted sine-Gordon theory and the minimal conformal series'},
Phys. Lett. B230 (1989) 103\toline{107}
%
\bibitem{ZZSmat}
A.B.\ Zamolodchikov and Al.B.\ Zamolodchikov,
{`Factorized S matrices in two dimensions as the exact solutions of  certain relativistic field models'},
Ann. of Phys. 120 (1979) 253\toline{291}
%
\bibitem{Sch}
K.\ Schoutens,
{`Supersymmetry and factorizable scattering'},
Nucl. Phys. B344 (1990) 665\toline{695}
%
\bibitem{yang}
C.N.\ Yang and C.P.\ Yang,
{`Thermodynamics of one-dimensional systems of bosons with repulsive delta function interaction'},
J. Math. Phys. 10 (1969) 1115\toline{1122}
%
\bibitem{ZamTBA}
Al.B.\ Zamolodchikov,
{`Thermodynamic Bethe Ansatz in relativistic models: Scaling 3-state Potts and Lee-Yang models'},
Nucl. Phys. B342 (1990) 695\toline{720}
%
\bibitem{FI}
P.\ Fendley and K.\ Intriligator,
{`Scattering and thermodynamics of fractionally charged supersymmetric solitons'},
Nucl. Phys. B372 (1992) 533\toline{558}
%
\bibitem{RTV1}
F.\ Ravanini, R.\ Tateo and A.\ Valleriani,
{`Dynkin TBAs'},
Int. J. Mod. Phys. A8 (1993) 1707\toline{1727}
%
\bibitem{RTV2}
F.\ Ravanini, R.\ Tateo and A.\ Valleriani,
{`A new family of diagonal A-D-E related scattering theories'},
Phys. Lett. B293 (1992) 361\toline{366}
%
\bibitem{FZa}
V.A.\ Fateev and Al.B.\ Zamolodchikov, 
{`Integrable perturbations of $\ZN$ parafermion models and the $O(3)$
  sigma model'},  
Phys. Lett. B271 (1991) 91\toline{100}
%
\bibitem{RTson}
R.\ Tateo,
{`The sine-Gordon model as $\frac{{\cal S}{\cal O}(n)_1 \times {\cal
S}{\cal O}(n)_1  }{ {\cal S}{\cal O}(n)_2} $ perturbed coset theory and generalizations'}, 
Int. J. Mod. Phys. A10 (1995) 1357\toline{1376}
%
\bibitem{RT}
R.\ Tateo,
{`New functional dilogarithm identities and  sine-Gordon Y-systems'},
Phys. Lett. B355 (1995) 157\toline{164}
%
\bibitem{Zab}
Al.B.\ Zamolodchikov, 
{`TBA equations for integrable perturbed $SU(2)_k \times SU(2)_l /  SU(2)_{k+l}$ coset models'}, 
Nucl. Phys. Lett. B366 (1991) 122\toline{132}  
%
\bibitem{AhnYL}
C.\ Ahn,
{`Thermodynamics and Form Factors of Supersymmetric Integrable Field theories'},
Nucl. Phys. B422 (1994) 449\toline{475}
%
\bibitem{FSn2}
P.\ Fendley and K.\ Intriligator,
{`Scattering and Thermodynamics in integrable $N=2$ theories'},
Nucl. Phys. B380 (1992) 265\toline{292}
%
\bibitem{MelzerSUSY}
E.\ Melzer,
{`Supersymmetric analogs of the Gordon-Andrews identities, and related TBA  systems'},
{\tt hep-th/9412154}
%
\bibitem{MS}
M.\ Moriconi and K.\ Schoutens,
{`Thermodynamic Bethe Ansatz for $N=1$ Supersymmetric Theories'},
Nucl. Phys. B464 (1996) 472\toline{491}
%
\bibitem{Zamy}
Al.B.\ Zamolodchikov,
{`On the thermodynamic Bethe ansatz equations for the reflectionless ADE scattering theories'},
Phys. Lett. B253 (1991) 391\toline{394}
%
\bibitem{KSS}
A.\ Kuniba, K.\ Sakai and J.\ Suzuki,
{`Continued fraction TBA and functional relations in XXZ model at root of unity'}, 
Nucl. Phys. B525 (1998) 597\toline{626}
%
\bibitem{DDV}
C.\ Destri and H.J.\ de Vega,
{`New thermodynamic Bethe ansatz equations without strings'},
Phys. Rev. Lett. 69 (1992) 2313\toline{2317}
%
\bibitem{FRT1}
G.\ Feverati, F.\ Ravanini and G.\ Takacs,
{`Nonlinear Integral Equation and Finite Volume Spectrum of
  Sine-Gordon Theory'}, 
Nucl. Phys. B540 (1999) 543\toline{586}
%
\bibitem{FRT4}
G.\ Feverati, F.\ Ravanini and G.\ Takacs,
{`Scaling functions in the odd charge sector of sine-Gordon/ massive 
Thirring theory'},
Phys. Lett. B444 (1998) 442\toline{450}
%
\bibitem{KBP}
A.\ Kl\"umper, M.T.\ Batchelor and P.A.\ Pearce,
{`Central charges of the 6- and 19-vertex models with twisted boundary conditions'},
J. Phys. A24 (1991) 3111\toline{3133}
%
\bibitem{PK}
P.A.\ Pearce and A.\ Kl\"umper, 
{`Finite-Size Corrections and Scaling Dimension of Solvable Lattice Models: An Analytic Method'},
Phys. Rev. Lett. 66 (1991) 974\toline{977}
%
\bibitem{Suzspin}
J.\ Suzuki,
{`Spinons in magnetic chains of arbitrary spins at finite temperatures'},
J. Phys. A32 (1999) 2341\toline{2359} 
%
\bibitem{Za}
A.B.\ Zamolodchikov, 
{`Renormalization group and perturbation theory about fixed points in two-dimensional field theory'},
Sov. J. Nucl. Phys. 46 (1987) 1090\toline{1096}
%
\bibitem{Zaint}
A.B.\ Zamolodchikov,
{`Integrable field theory from conformal field theory'}, 
Adv. Stud. in Pure Maths. 19 (1989) 641\toline{674}
%
\bibitem{ZRSOS}
Al.B.\ Zamolodchikov, 
{`Thermodynamic Bethe Ansatz for RSOS scattering theories'},
Nucl. Phys. B358 (1991) 497\toline{523}
%
\bibitem{Zc}
Al.B.\ Zamolodchikov, 
{`From tricritical Ising to critical Ising by Thermodynamic Bethe Ansatz'},
Nucl. Phys. B358 (1991) 524\toline{546}
%
\bibitem{Zd}
Al.B.\ Zamolodchikov, 
{`Thermodynamics of imaginary coupled  sine-Gordon: Dense polymer finite-size scaling function'}, 
Phys. Lett. B335 (1994) 436\toline{443}
%
\bibitem{Zamp}
Al.B.\ Zamolodchikov, 
{`Painleve\'e III and 2D Polymers'},
Nucl. Phys. B432 (1994) 427\toline{456}
%
\bibitem{FMQR}
D.\ Fioravanti, A.\ Mariottini, E.\ Quattrini and F.\ Ravanini,
{`Excited state Destri-De Vega equation for sine-Gordon and restricted sine-Gordon models'}, 
Phys. Lett. B390 (1997) 243\toline{251}
%
\bibitem{FRT3}
G.\ Feverati, F.\ Ravanini and  G.\ Takacs,
{`Nonlinear integral equation and finite volume spectrum of minimal models perturbed by $\phi_{1,3}$'}, 
Nucl. Phys. B570 (2000) 615\toline{643},
%
\bibitem{DDT}
P.\ Dorey, C.\ Dunning and R.\ Tateo,
{`New families of flows between two-dimensional conformal field theories'},
Nucl. Phys. B578 (2000) 699\toline{727}
%
\bibitem{KM1}
T.R.\ Klassen and E.\ Melzer,
{`Purely elastic scattering theories and their ultraviolet limits'},
Nucl. Phys. B338 (1990) 485\toline{528}
%
\bibitem{Kir}
A.N.\ Kirillov,
{`Dilogarithm identities'},
Prog. Theor. Phys. Suppl. 118 (1995) 61\toline{142}
%
\bibitem{DTc}
P.\ Dorey and R.\ Tateo,
{`Differential equations and integrable models: the $SU(3)$ case'},
Nucl. Phys. B571 (2000) 583\toline{606}
%
\bibitem{PF2}
P.\ Fendley,
{`Integrable sigma models and perturbed coset models'},
JHEP 0105 (2001) 050
%
\bibitem{KU1}
 K.\ Kobayashi and T.\  Uematsu,
{`$N=2$ supersymmetric sine-Gordon theory and conservation laws'},
Phys. Lett. B264 (1991) 107\toline{13}
%
\bibitem{MM}
J.\ Maldacena and L.\  Maoz,
{`Strings on pp-waves and massive two dimensional field theories'},
{\tt hep-th/0207284}
%
\bibitem{War}
N.P.\ Warner,
{`Supersymmetry in boundary integrable quantum field theories'},
Nucl. Phys. B450 (1995) 663\toline{694}
%
\bibitem{Nep}
R.I.\ Nepomechie,
{`Boundary S matrices with $N=2$ supersymmetry'},
Phys. Lett. B516 (2001) 161\toline{164}
;
{`The boundary $N=2$ supersymmetric sine-Gordon model'},
Phys. Lett. B516 (2001) 376\toline{382}
%
\bibitem{KU2}
K.\ Kobayashi and T.\  Uematsu,
{`S matrix of $N=2$ Supersymmetric Sine-Gordon model'},
Phys. Lett. B275 (1992) 361\toline{370}
;
{`Quantum conserved charges and S matrices in $N=2$ supersymmetric sine-Gordon theory'},
Prog. Theor. Phys. Suppl. 110 (1992) 347\toline{364}
%
\bibitem{FMVW}
P.\ Fendley, S.D.\ Mathur, C.\ Vafa and N.P.\ Warner,
{`Integrable deformations and scattering matrices for the $N=2$ supersymmetric discrete series '},  
Phys. Lett. B243 (1990) 257\toline{264}
%
\bibitem{Fatss}
V.A.\ Fateev,
{`The sigma model (dual) representation for a two-parameter family of integrable quantum field theories'},
Nucl. Phys. B473 (1996) 509\toline{538}
%
\bibitem{FOZ}
V.A.\ Fateev, E.\ Onofri  and Al.B.\ Zamolodchikov, 
{`Integrable deformations of the $O(3)$ sigma model. The sausage model'},
Nucl. Phys. B406 (1993) 521\toline{565}
%
\bibitem{F}
V.A.\ Fateev,
{`Integrable deformations in $Z_n$-symmetrical models of the conformal
  quantum field theory'},
Int. J. Mod. Phys. A6 (1991) 2109\toline{2132}
%
\bibitem{FB}
P.\ Baseilhac and V.A.\ Fateev,
{`Expectation values of local fields for a two-parameter family of integrable models and related perturbed conformal field theories'},
Nucl. Phys. B532 (1998) 567\toline{587}
%
\bibitem{TCDw}
C.\ Dunning,
{`Massless flows between minimal $W$ models'},
 Phys. Lett. B537 (2002) 297\toline{305}



\end{thebibliography}
\end{document}